# Microwave Imaging of Edge Conductivity in Graphene at Charge Neutrality and Quantum Hall States

Hongtao Yan[1†], Chun-Chih Tseng[2†], Anzhuoer Li[1†], Manish Kumar[2], Kaile Wang[1], Shizai Chu[1], Kenji Watanabe[3], Takashi Taniguchi[4], Allan H. MacDonald[1*], Matthew Yankowitz[2*], Keji Lai[1*]

[1] Department of Physics, The University of Texas at Austin, Austin, Texas 78712, USA
[2] Department of Physics, University of Washington, Seattle, WA 98195, USA
[3] Research Center for Electronic and Optical Materials, National Institute for Materials Science, 1-1 Namiki, Tsukuba 305-0044, Japan
[4] Research Center for Materials Nanoarchitectonics, National Institute for Materials Science, 1-1 Namiki, Tsukuba 305-0044, Japan

[†] These authors contributed equally to this work.

[*] Corresponding authors

## Abstract

We report local conductivity imaging of edge states in monolayer graphene by millikelvin microwave impedance microscopy (MIM). At the charge-neutrality point, as the magnetic field increases, the local conductivity at the edge drops to zero more slowly than in the bulk. This behavior is consistent with the calculated spatial profile of the charge gap in the canted antiferromagnetic phase. For comparison, we also perform microwave imaging of integer quantum Hall states away from neutrality, which host dissipationless chiral edge channels. The evolution of the edge signal as a function of the bulk gap is fundamentally different between the Landau level filling factor $\nu = 0$ and $|\nu| \geq 1$ integer quantum Hall states, which can be qualitatively explained by numerical simulations and theoretical analysis. Our results provide a comprehensive microscopic picture of the edge and bulk states as the Fermi level moves across the unique Landau-level spectrum of graphene.

Under a strong perpendicular magnetic field ($B_\perp$), the low-energy spectrum of Dirac fermions in monolayer graphene is quantized into a set of relativistic Landau levels (LLs) with energy $\epsilon_N = \text{sgn}(N)v_F\sqrt{2\hbar e B_\perp |N|}$, where $v_F$ is the Fermi velocity and $N$ is the LL index[1-12]. Of special interest is the zero-energy LL ($N = 0$), in which the electron wavefunction in one valley is localized on a single sublattice. To describe the $N = 0$ LL, it is convenient to use one isospin space with SU(4) symmetry that incorporates valley (or sublattice) and spin degrees of freedom in the framework of quantum Hall ferromagnetism[13,14] (QHFM). The four-fold flavor degeneracy can be lifted by electron-electron interactions under a strong $B$-field, resulting in many-body gaps that are larger than typical single-particle Zeeman or valley gaps[4,14-17]. Specifically, at the charge neutrality point (CNP) (i.e., $\nu = 0$)[6,10,18-20], the competition between short-range anisotropic interaction and Zeeman field can lead to four possible broken-symmetry QHFM states[21-29]: fully spin-polarized ferromagnetic (FM) phase, sublattice-polarized but spin-unpolarized charge density wave (CDW) phase, canted antiferromagnetic (CAF) phase with noncollinear spin polarizations, and spin-unpolarized intervalley coherent (IVC) state with Kekulé lattice distortion. The CNP in monolayer graphene thus provides a fertile playground for studying correlated quantum phenomena in two-dimensional (2D) systems.

The spatial arrangement of charges at the CNP in graphene has been studied by atomically resolved scanning tunneling microscopy (STM) experiments[30-32]. On the other hand, the local conductivity of these QHFM phases has not been fully explored. Theoretically, the FM phase at $\nu = 0$ is characterized by an insulating bulk and conductive quantum spin Hall edge[22-28], which is supported by transport data under tilted $B$-fields[10]. Both the bulk and edge of the other three phases are expected to be insulating, with subtle differences in the spatial distribution of excitation gaps[22-25,28]. For CDW and IVC phases, the local energy gap increases monotonically from the bulk to edge of the sample[23-25,28]. In contrast, for the CAF phase, the charge gap in the bulk first reduces when approaching the edge and then diverges at the physical boundary[22,24,25,28]. The spatial evolution of bulk and edge conductivity is therefore critical for the understanding of broken-symmetry quantum phases in graphene.

In this work, we perform spatially resolved conductivity imaging in monolayer graphene by microwave impedance microscopy[33,34]. All integer quantum Hall effect (IQHE) states in the first few LLs are resolved by MIM point spectroscopy as a function of gate voltage. At $\nu = 0$, as the perpendicular *B*-field increases, the bulk conductivity drops to zero much faster than that of the edge, consistent with the expected behavior of the CAF phase. For comparison, we also map out the local conductivity distribution of $|\nu| \geq 1$ IQHE states, which have drastically different energy gaps. The MIM signals of edge-like features at the CNP under intermediate *B*-fields are qualitatively different from those of topological chiral edges at other IQHE states, which can be explained by theoretical analysis and numerical simulations.

Fig. 1a shows schematics of our device structure and experimental setup. The monolayer graphene is encapsulated by top and bottom hexagonal boron nitride (hBN). Few-layer graphite (FLG) is used as the bottom gate. Three devices have been studied with consistent results (Supplementary Information S1). The transport and microscopy experiments are performed at $T = $ 180 mK in a dilution refrigerator (DR) platform[35]. For MIM measurements, an etched tungsten tip is mounted on a quartz tuning fork (TF)[36,37] for topographic feedback. After locating the device, we lift the tip by 30 nm and perform MIM imaging in the constant-height mode to minimize perturbation to the sample. A microwave signal with frequency $f_{MIM}$ = 2.86 GHz and input power 0.1 ~ 1 μW is fed to the tip, and the reflected signal is amplified and demodulated by the MIM electronics. The output baseband signals, which are proportional to the imaginary (MIM-Im) and real (MIM-Re) parts of the tip-sample admittance, are further demodulated at the TF resonant frequency $f_{TF}$ of ~ 32 kHz. Using finite-element analysis (FEA)[38], we can simulate MIM signals as a function of 2D sheet conductance $\sigma_{2D}$ of a uniform graphene layer, as shown in Fig. 1b. Note that $f_{MIM}$ is much lower than the first edge magnetoplasma (EMP) resonant frequency[39] such that the lumped-element model is sufficient for simulations. The signal in the capacitive channel (MIM-Im) increases monotonically with respect to $\sigma_{2D}$, whereas the signal in the loss channel (MIM-Re) displays a peak at intermediate $\sigma_{2D}$. In this work, since the local conductivity can vary on a length scale smaller than the tip diameter (~ 300 nm), more sophisticated simulations are needed for interpreting the MIM data (to be detailed below).

Fig. 1c shows the gate-dependent MIM signals taken at $B$ = 8 T when the tip oscillates at a single point above the bulk of graphene. All IQHE states from the $N$ = 0 and ±1 LLs are resolved as local minima in MIM-Im, whose strengths agree with the relative sizes of energy gaps in graphene. As schematically shown in Fig. 1d, the strongest states at $\nu$ = ±2 and ±6 correspond to cyclotron gaps. Due to the unique property of exact valley-sublattice correspondence in the zeroth

LL, the symmetry-broken gaps at $\nu = 0$ and $\pm 1$ are nearly equal and larger than spin and valley gaps at higher LLs. For $|N| \geq 1$ LLs, the Zeeman splitting ($\nu = \pm 4, \pm 8 \ldots$) is larger than the valley splitting (odd fillings except for $\pm 1$). Such a hierarchy of gap sizes in graphene is well documented by transport studies[4,15]. Fig. 1e shows the gate voltages at the local minima $V_{min}$ as a function of filling factors, which coincide with values from the transport data (Supplementary Information S2). Thus, we do not observe the mismatch between transport plateaus and bulk fillings reported by an earlier MIM work[40], which attributes the discrepancy to the presence of a wide edge strip with charge accumulation[41]. Since our gate dielectric is much thinner (50 nm) than that (> 300 nm) in Ref. [40], the effect proposed in that work is likely negligible in the current study.

The gate voltage $V_G$ corresponding to $\nu = 0$ in this device is carefully determined in the experiment (Supplementary Information S3). We then fix $V_G$ at the CNP and take $B$-dependent MIM images across the sample (dashed line in Fig. 1a), as displayed in Figs. 2a and 2c. In the constant-height mode, there is a background tip-sample capacitance difference between the etched and unetched regions of the sample (Fig. 1a). We note that the MIM-Im signals on graphene are highest (lowest) at 0 T (8 T), whereas the MIM-Re signals are essentially zero at both 0 T and 8 T. The vanishing MIM-Re indicates that the near-field tip-sample interaction is lossless or purely capacitive, which could take place in two regimes. First, the graphene is in the conductive limit ($\sigma_{2D} > 10^{-3}$ S·sq) such that the GHz electric fields are fully screened by the 2D layer, corresponding to the situation at $B = 0$ T. Second, the graphene is in the resistive limit ($\sigma_{2D} < 10^{-9}$ S·sq) such that the GHz electric fields completely transmit through the 2D layer, corresponding to the situation at $B = 8$ T. As a result, the background MIM-Im signals can be removed by subtracting a single line taken at $B = 8$ T from the raw data (included in Supplementary Information S4). We will justify the background subtraction in the simulation below. The resulting MIM-$\Delta$Im images are shown in Fig. 2b, with the corresponding average line profiles plotted in Fig. 2d. As $B$ increases from zero, the bulk MIM-$\Delta$Im signals drop rapidly and saturate to zero beyond 4 T. Near the edge of graphene, the signals drop more slowly than the bulk with increasing $B$-fields. Using the simulated MIM response curves in Fig. 1b, we can quantitatively extract the local conductivity profiles (Supplementary Information S5), which clearly show that finite conductivity persists at the graphene edge up to 6 T.

The conductive edge features at the CNP under intermediate $B$-fields are fundamentally different from the topological boundaries in $|\nu| \geq 1$ IQHE states. Fig. 3 shows the $V_G$-dependent MIM signals across the sample. For the Im channel, we again display MIM-$\Delta$Im images by removing the background at $\nu = 0$ and $B = 8$ T (raw data in Supplementary Information S4). The entrance into and departure from each IQHE state by gate tuning are vividly seen in the MIM data. Moreover, as plotted in Fig. 3d, the spatial distribution of MIM signals at integer fillings exhibits a systematic dependence on the energy gaps. For main-sequence IQHE states with cyclotron gaps ($\nu = 2$ and 6), small MIM-$\Delta$Im peaks occur at the edge, whereas MIM-Re vanishes everywhere on the sample. As the gap shrinks, the MIM-$\Delta$Im edge peaks grow in height and width, and merge with the bulk for $\nu = 3$ and 4 states that are not fully insulating. In the meantime, the MIM-Re channel also develops peaks near the edge, but with positions that do not overlap with the corresponding MIM-$\Delta$Im peaks. This is in sharp contrast to the scenario at $\nu = 0$ under intermediate $B$-fields (Fig. 2), where peak features in both channels always align with each other.

The different behaviors in Figs. 2 and 3 suggest that edge-like features in the MIM data do not share the same underlying physics at $\nu = 0$ versus $|\nu| \geq 1$. A key distinction is that the $\nu = 0$ state is expected to be topologically trivial in nature, whereas all other IQHE states are nontrivial. Following the theoretical framework in Ref. [23], we perform Hartree-Fock calculations (Supplementary Information S6) to analyze the spatial distribution of excitation gaps near the graphene edge for the three insulating QHFM phases at the CNP. For the CDW and IVC phases, the charge gap at $\nu = 0$ increases monotonically towards the edge. In contrast, for the CAF phase, the calculated local bandgap first decreases when approaching the edge before diverging at the physical boundary. We then compute the local conductivity profiles of the three phases (Supplementary Information S6). Within several magnetic lengths $\left(l_B = \sqrt{\hbar/eB}\right)$ near the edge, the CAF phase exhibits higher conductivity in this gap-narrowing region than that in the bulk, which does not occur for the CDW and IVC phases. In the 3D FEA simulation (Supplementary Information S7), we assume a spatially constant edge conductivity $\sigma_e > \sigma_b$ (bulk conductivity) for this region. Using $\sigma_e$ and $\sigma_b$ values on the order of $0.01 \sim 0.1$ $e^2/h$, which are physically reasonable due to thermal activation, we can qualitatively reproduce the field-dependent MIM line profiles in the simulation, as shown in Fig. 4a. More importantly, after removing the background MIM-Im signals, the simulated peaks in both MIM-ΔIm and MIM-Re channels occur at the same location under intermediate $B$-fields, again consistent with the observation in Fig. 2d.

In contrast, the topologically nontrivial $|\nu| \geq 1$ IQHE states are characterized by the presence of chiral edge channels. For simplicity, we model the dissipationless edge as a perfectly conductive ($\sigma_{xx} \to \infty$) strip in Fig. 4b. The bulk of graphene is described by a 2D conductivity tensor with the diagonal components $\sigma_{xx} = \sigma_{yy} = \sigma_b$ and off-diagonal $\sigma_{xy} = -\sigma_{yx} = \nu e^2/h$. The transition region between the edge and bulk is modeled as an inner strip with finite conductivity $\sigma_{xx} = \sigma_e$. By adjusting the values of $\sigma_e$ and $\sigma_b$ and ensuring they remain physically reasonable (Supplementary Information S7), we can simulate a series of MIM line profiles as the bulk gap increases, which capture key features in the experimental data (Fig. 3d). In this case, the MIM-ΔIm signals are dominated by the dissipationless edge, so the corresponding peaks are at or very close to the physical boundary. The MIM-Re signals, if exist, originate from the transition region and/or the bulk. Consequently, the MIM-Re peaks shift inward with respect to the MIM-ΔIm counterparts, consistent with the experimental observation. This offset between peak positions in the two channels can be seen in other MIM studies of topologically nontrivial systems such as quantum Hall[40,42], quantum spin Hall[43,44], and quantum anomalous Hall effects[35,45].

In the following, we discuss the interpretation of our results. First, it is unlikely that the MIM edge signals at the CNP under intermediate $B$-fields can be attributed to extrinsic effects such as etching-induced charge defects, which are localized within a few lattice constants near the physical boundary. In order to reproduce the experimental data in the FEA simulation (Supplementary Information S7), one has to assume a high $\sigma_e$ well into the metallic regime ($\sim 10$ $e^2/h$) within a 1-nm strip, which is physically unreachable for defective states. Second, the measured IQHE edge width at $\nu = 2$ ($\sim 600$ nm) is much larger than typical compressible strips, which are expected to be on the order of $l_B$ ($\sim 10$ nm at 8 T). We attribute this phenomenon to the combined effect of relatively blunt tip (300 ~ 400 nm in diameter), large tip oscillation amplitude (120 nm peak-to-peak), and lift-mode operation (30 nm above graphene). In fact, using a sharper tip and smaller tip-sample distance, we measure a much narrower width of ~ 160 nm at $\nu = 2$ on a different device with unetched natural graphene edges (Supplementary Information S1). The apparent edge width is thus limited by the spatial resolution and does not reflect its actual size.

Finally, recent STM studies show that CAF and IVC phases can both exist at the CNP but under different dielectric environment[31,32]. Specifically, the CAF phase occurs when the graphene is strongly screened by the $SrTiO_3$ substrate, whereas the IVC phase is seen when the graphene is placed on $hBN/SiO_2$ with weak screening[31]. We note that the upper surface of graphene is kept in vacuum for the STM experiments, whereas both sides of our graphene device are in contact with hBN. It is thus feasible that the additional screening from the top hBN favors the CAF phase in our sample. Further experimental and theoretical studies are needed to elucidate the competition between various QHFM phases. Moreover, CAF quantum Hall phases have also been predicted and experimentally studied in GaAs bilayer quantum wells[46]. Extending the MIM experiment to such systems would be an exciting direction for future works.

In summary, we perform local conductivity imaging of gated graphene devices by millikelvin MIM. At the charge neutrality point, finite conductivity appears at the edge of graphene under intermediate magnetic fields where the bulk is fully insulating. The phenomenon suggests that the charge excitation gap is smaller near the edge compared to the bulk, consistent with the canted antiferromagnetic ground-state configuration. For comparison, the local conductivity distribution across the sample is mapped out at several $|\nu| \geq 1$ integer quantum Hall states. The relative peak positions in the two MIM channels are different between topologically trivial ($\nu = 0$) and non-trivial ($|\nu| \geq 1$ IQHE) edge states, which can be reproduced by finite-element analysis. Our results are important for the understanding of correlated quantum phases in graphene.

**Acknowledgement**: The MIM work at UT-Austin was supported by the U.S. Department of Energy (DOE), Office of Science, Basic Energy Sciences, under Award DE-SC0019025. Sample fabrication and initial characterization work at the University of Washington was supported by the U.S. Department of Energy (DOE), Office of Science, Basic Energy Sciences, under Award DE-SC0023062. The numerical simulation was supported by the Gordon and Betty Moore Foundation, grant GBMF12238 (https://doi.org/10.37807/gbmf12238), and the Welch Foundation grant F-1814. The theoretical analysis was supported by the NSF through the Center for Dynamics and Control of Materials, an NSF Materials Research Science and Engineering Center (MRSEC) under Cooperative Agreement DMR-2308817.

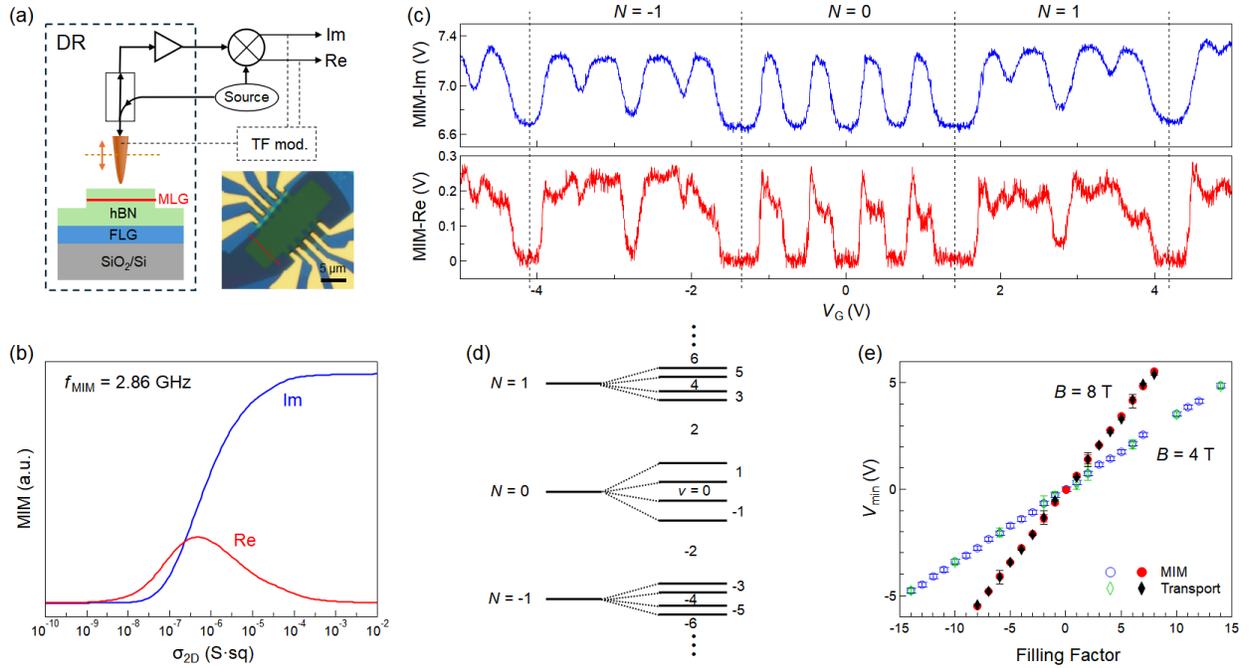

**FIG. 1.** (Color online) **(a)** Schematics of the experimental setup and device structure, with an optical image of the device in the bottom right. **(b)** Simulated MIM signals as a function of the 2D sheet conductance of a uniform graphene layer at $f_{MIM}$ = 2.86 GHz. **(c)** Gate-dependent MIM signals when the tip is positioned at a single point above the graphene bulk, measured at $T$ = 180 mK and $B$ = 8 T. **(d)** Schematics of energy levels in graphene under a strong perpendicular $B$-field. The gaps at each filling factor are labeled in the graph. **(e)** Gate voltages at local minima in the MIM-Im and transport data as a function of filling factors.

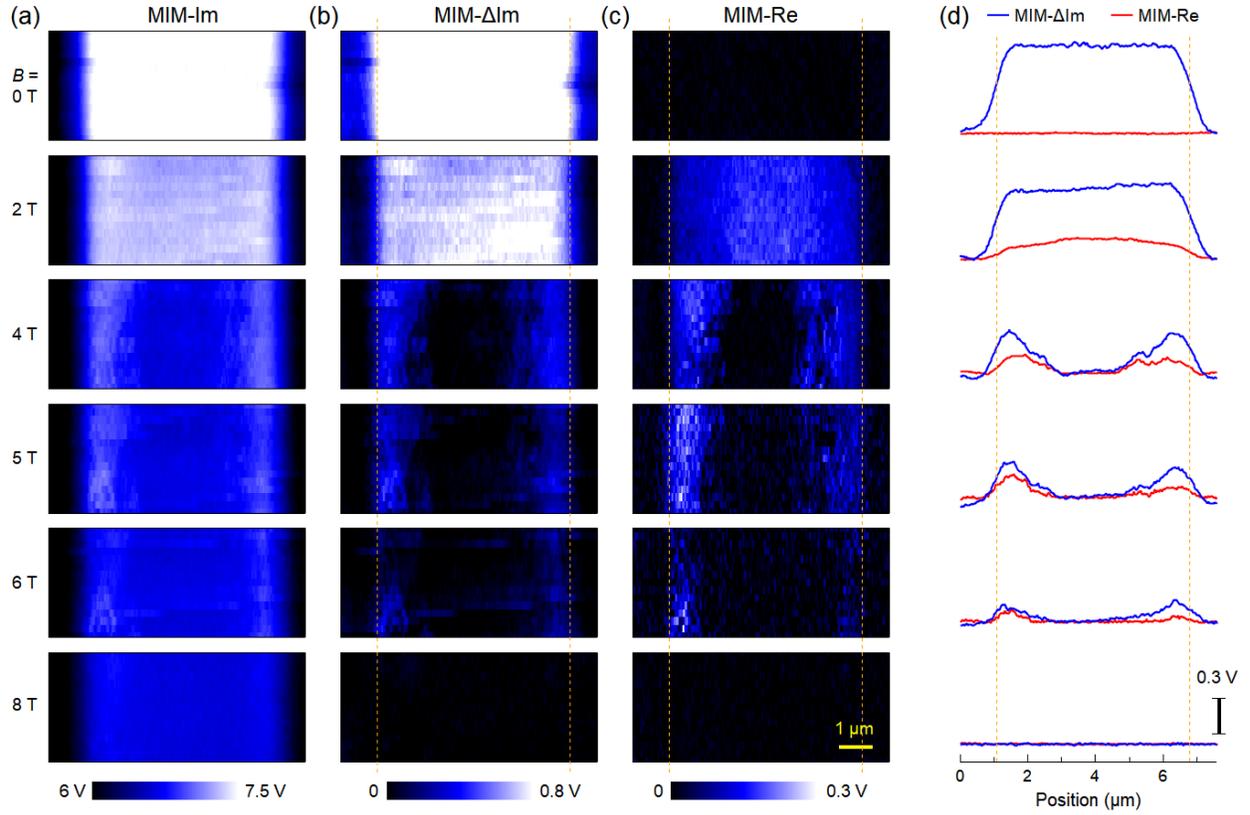

**FIG. 2.** (Color online) **(a)** Field-dependent MIM-Im, **(b)** MIM-ΔIm, and **(c)** MIM-Re images taken near the dashed line in the inset of Fig. 1a and at the CNP. Each image spans 1.4 μm along the vertical direction. **(d)** Corresponding average MIM-ΔIm and MIM-Re line profiles under each $B$-field. Vertical dashed lines in **b** – **d** indicate the etched physical edges.

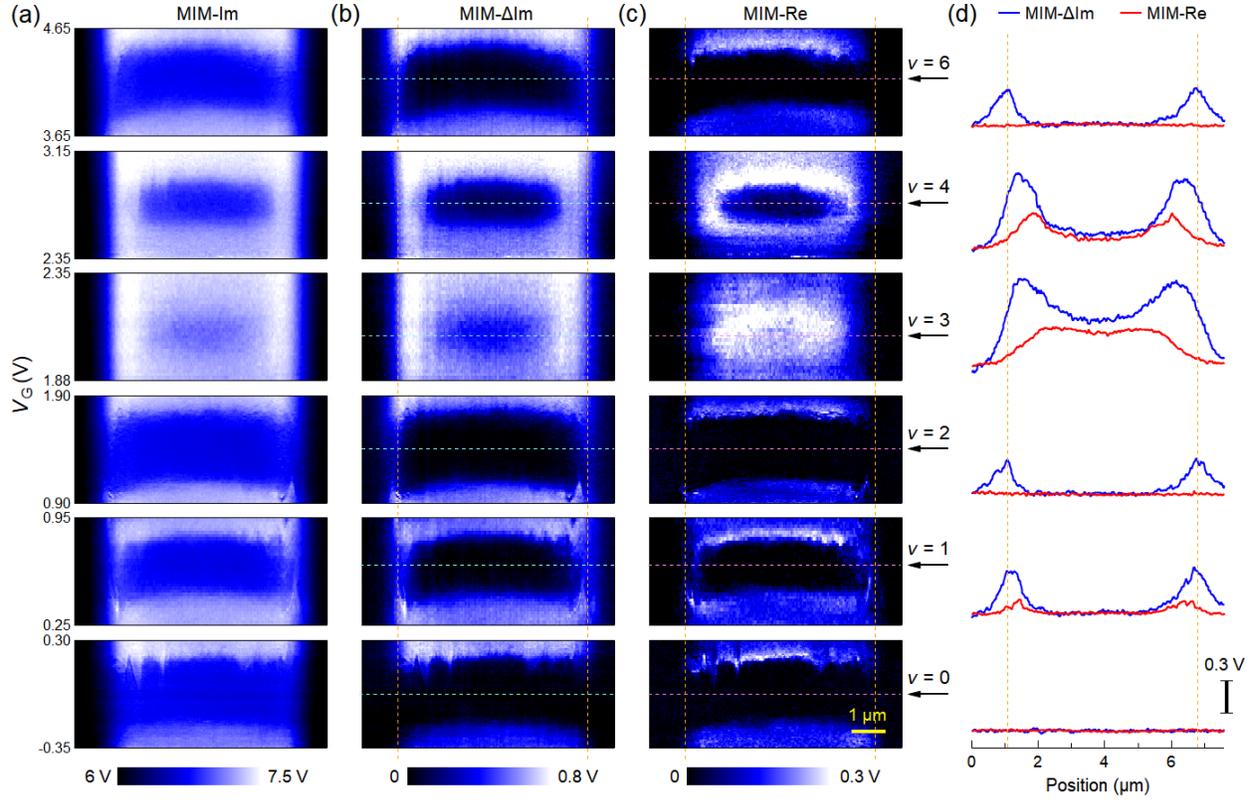

**FIG. 3.** (Color online) **(a)** Gate-dependent MIM-Im, **(b)** MIM-ΔIm, and **(c)** MIM-Re data across the sample (dashed line in the inset of Fig. 1a) around various IQHE states and the CNP at 8 T. Note that the y-axis is the gate voltage instead of position. **(d)** Corresponding MIM-ΔIm and MIM-Re line profiles at exact integer fillings. Vertical dashed lines in **b** – **d** indicate the etched physical edges.

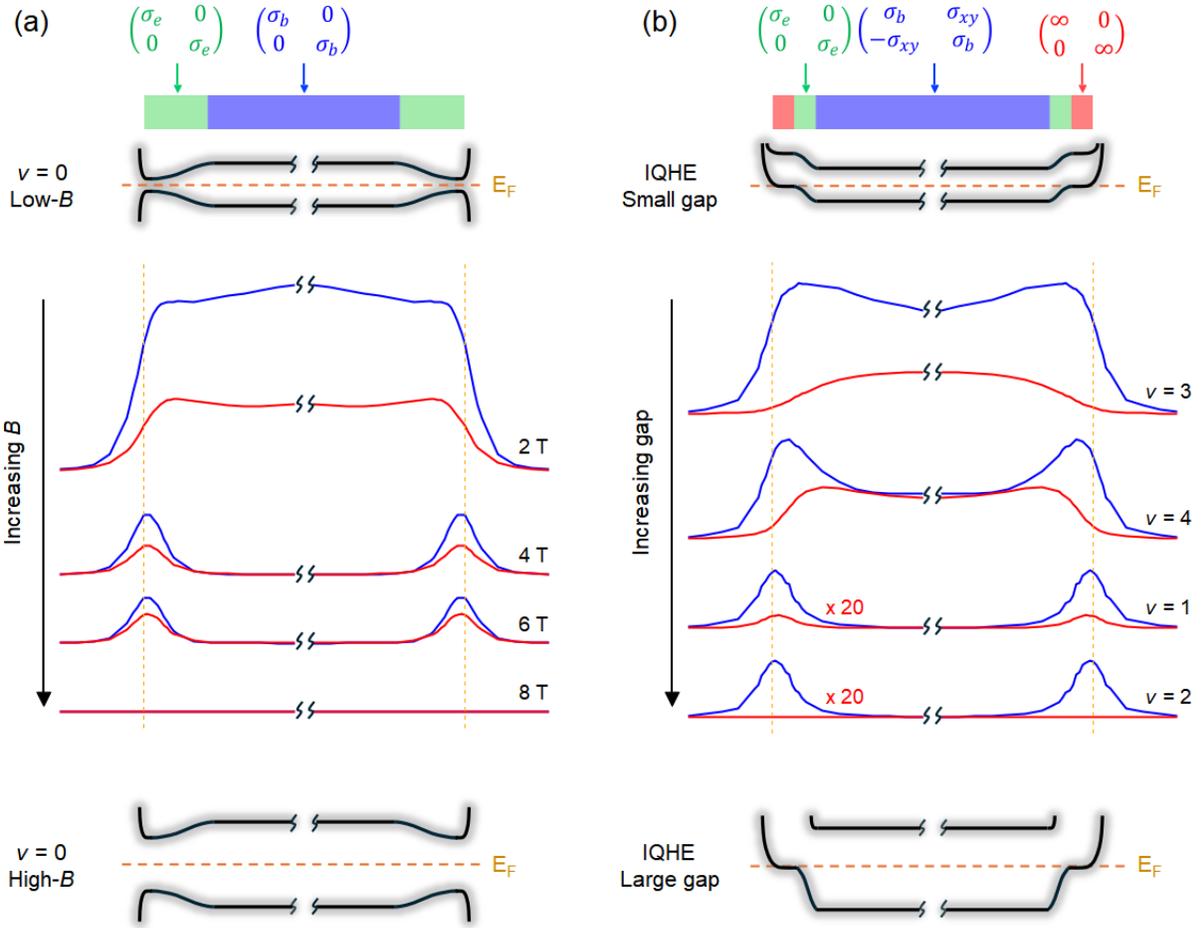

**FIG. 4.** (Color online) **(a)** Simulated MIM signals at the CNP with increasing *B*-fields and **(b)** at various IQHE states with increasing energy gaps. The schematics on the top illustrate the FEA model with conductivity tensors for the bulk and edge regions, as well as the band diagrams of the CAF phase under a low *B*-field in **a** and IQHE state with a small energy gap in **b**. The shades of each energy level represent the disorder broadening. The MIM-ΔIm and MIM-Re line profiles as increasing *B*-fields are simulated by selecting $\sigma_e$ and $\sigma_b$ to qualitatively match the experimental data. The schematics on the bottom illustrate the band diagrams of the CAF phase under a high *B*-field in **a** and the IQHE state with a large energy gap in **b**. Vertical dashed lines indicate the etched physical edges.

*Supplementary Information for*

# Microwave Imaging of Edge Conductivity in Graphene at Charge Neutrality and Quantum Hall States

Hongtao Yan[1†], Chun-Chih Tseng[2†], Anzhuoer Li[1†], Manish Kumar[2], Kaile Wang[1], Shizai Chu[1], Kenji Watanabe[3], Takashi Taniguchi[4], Allan H. MacDonald[1*], Matthew Yankowitz[2*], Keji Lai[1*]

[1]Department of Physics, The University of Texas at Austin, Austin, Texas 78712, USA
[2]Department of Physics, University of Washington, Seattle, WA 98195, USA
[3]Research Center for Electronic and Optical Materials, National Institute for Materials Science, 1-1 Namiki, Tsukuba 305-0044, Japan
[4]Research Center for Materials Nanoarchitectonics, National Institute for Materials Science, 1-1 Namiki, Tsukuba 305-0044, Japan

[†] These authors contributed equally to this work.

[*] Corresponding authors

**Section S1.  Repeatability of microwave impedance microscopy (MIM) results on multiple graphene devices**

We measured three different monolayer graphene devices, labeled as Device #1, #2, and #3. All data presented in the main text were obtained from Device #1. All MIM measurements were carried out in the constant-height mode with an oscillation amplitude of 60 nm. The minimum tip-sample spacing was 30 nm for Device #1 and #2, and 20 nm for Device #3.

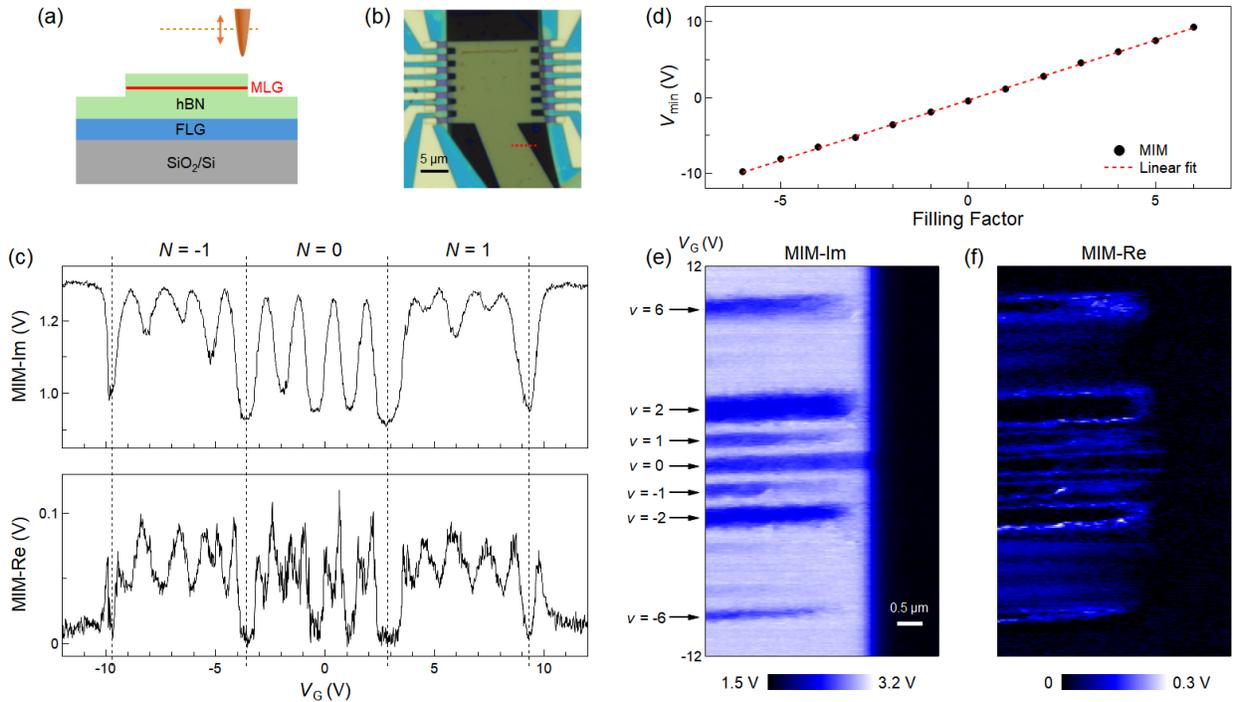

**FIG. S1.** **(a)** Schematic and **(b)** optical image of the graphene Device #2. **(c)** Gate-dependent MIM-Im and MIM-Re signals when the tip is positioned at a single point above the bulk, taken at $T = 130$ mK and $B = 8$ T. **(d)** Gate voltages at local minima in the MIM-Im data as a function of filling factors. The red dashed line is the linear fit. **(e)** Gate-dependent MIM-Im and **(f)** MIM-Re images across a single line (dashed line in **b**). The bulk filling factors are labeled in **e**.

The schematic and optical image of Device #2 are shown in Figs. S1a and S1b, respectively. The monolayer graphene is encapsulated by top (5.6 nm) and bottom (106 nm) hexagonal boron nitride (hBN), with a layer of 10 nm few-layer graphite (FLG) as the bottom gate. The height of the etched step edge is 20 nm. The gate-dependent imaginary (MIM-Im) and real parts (MIM-Re) of the MIM signal on the graphene bulk measured at $T = 130$ mK and $B = 8$ T are displayed in Fig. S1c. The local minima in the MIM-Im data correspond to the Landau level filling factor $\nu = 0$ (charge neutrality point, CNP) and $|\nu| \geq 1$ integer quantum Hall effect (IQHE) states. The gate voltages at these local minima are plotted as a function of filling factors in Fig. S1d. The MIM signals along a single line across the sample edge (dashed line in Fig. S1b) during the gate sweep are shown as false-color maps in Fig. S1e and Fig. S1f. At the CNP, the sample is insulating across the bulk and edge, as indicated by the low MIM-Im and vanishing MIM-Re signals. This is in sharp contrast to the IQHE states at, e.g., $\nu = \pm 1$ and $\pm 2$, where the edge remains conductive from the MIM data. In short, the MIM results of Device #2 are consistent with Device #1 presented in the main text.

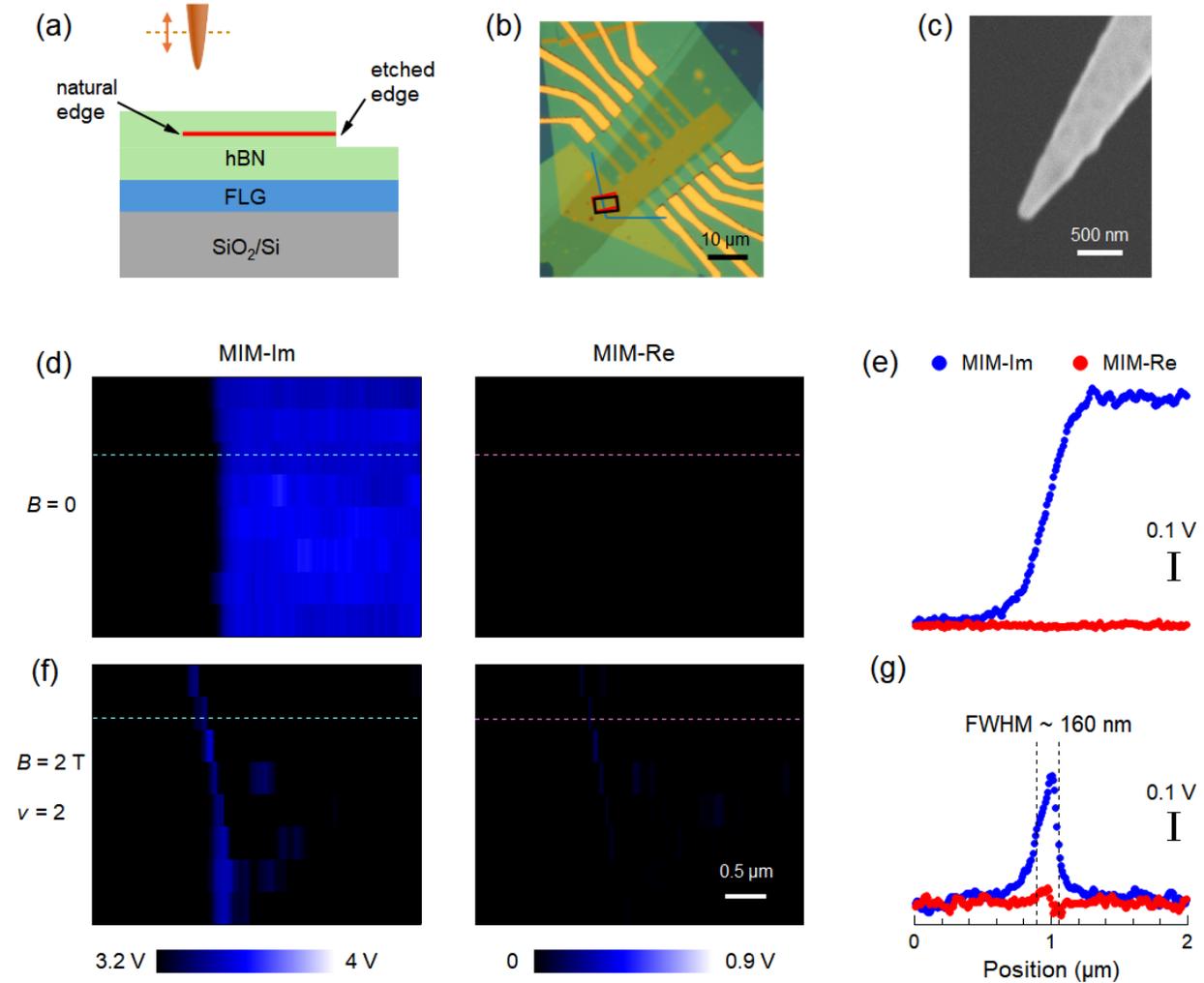

**FIG. S2.** (a) Schematic and (b) optical image of the graphene Device #3. The blue lines indicate unetched natural graphene edges. (c) Scanning electron microscopy image of the MIM tip for this experiment. (d) MIM images at $V_G = 0$ V and $B = 0$ T, taken in the red box in **b**. (e) MIM line profiles across the dashed line in **d**. (f) MIM images at $V_G = 0.2$ V and $B = 2$ T ($\nu = 2$), taken in the box in **b**. (g) MIM line profiles across the dashed line in **f**, showing an edge peak with the FWHM of ~160 nm.

The schematic and optical image of Device #3 are shown in Figs. S2a and S2b, respectively. This lower-quality device has more imperfections such as interlayer bubbles than the other two devices. As a result, we did not perform extensive experiments on this sample. Instead, we focused on the apparent width of the IQHE edge states by using a sharp tip (Fig. S2c). In addition, only part of the graphene is etched in Device #3, leaving segments of natural edges for MIM imaging (see schematic in Fig. S2a and image in Fig. S2b). As a result, no background subtraction is needed for this device. Figs. S2d and S2f show the MIM images at $B = 0$ T and $B = 2$ T ($\nu = 2$ IQHE state), respectively, at $T = 180$ mK. The entire graphene sample is at the conductive limit ($\sigma_{2D} > 10^{-3}$ S·sq) at $B = 0$ T, as indicated by the high MIM-Im and zero MIM-Re signals (Fig. S2e). On the other hand, the $\nu = 2$ IQHE state is characterized by the presence of a conductive edge and an insulating bulk based on the MIM signals (Fig. S2g). Note that the full width at half maximum (FWHM) of the edge feature in MIM-Im line profile at $\nu = 2$ is ~ 160 nm, which is presumably due to the combined effect of tip diameter (~ 100 nm), tuning fork oscillation (60 nm in amplitude) and constant-height mode (20 nm lift height). In all, we believe that the apparent peak width of ~ 600 nm in the MIM data on Device #1 presented in the main text is due to the spatial resolution rather than the actual edge width.

## Section S2. Transport and MIM data

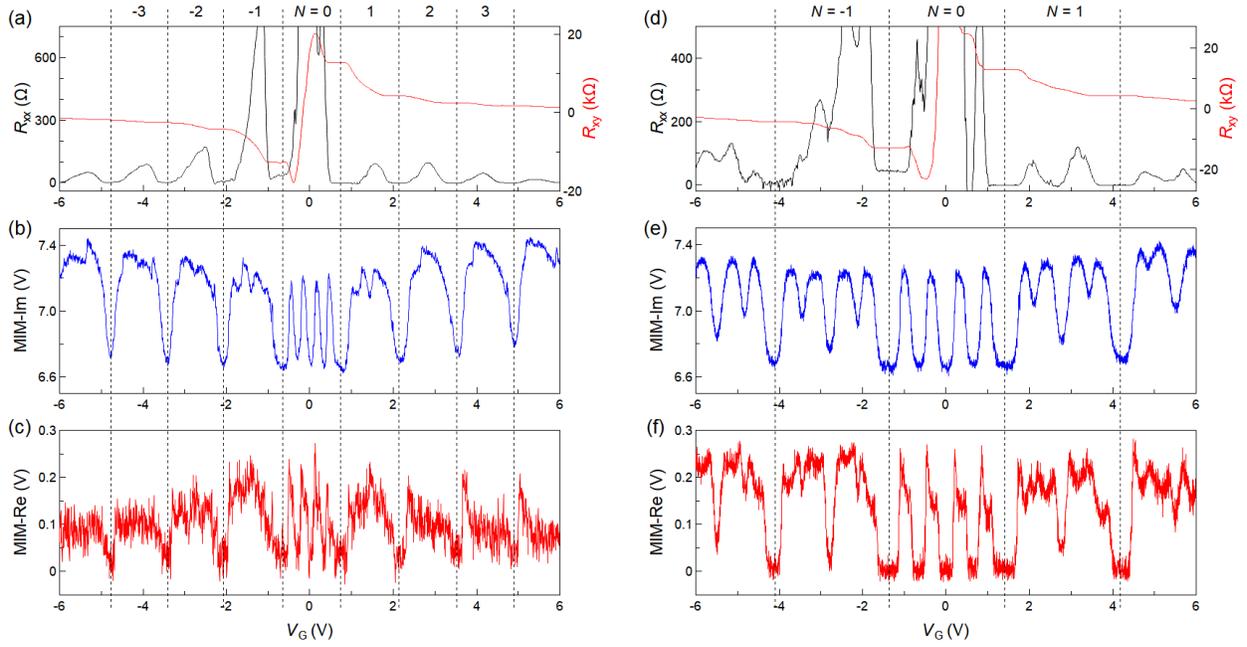

**FIG. S3.** (a) Longitudinal resistance $R_{xx}$ and transverse resistance $R_{xy}$ of Device #1 measured at $T = 180$ mK and $B = 4$ T. (b) Corresponding gate-dependent bulk MIM-Im and (c) MIM-Re signals at $B = 4$ T. (d-f) Same as a-c but at $B = 8$ T.

The transport and MIM data of Device #1 at $T = 180$ mK and $B = 4$ T and 8 T are shown in Fig. S3. The Ohmic contacts of this device (thus the transport data) are not of the best quality. Nevertheless, the most prominent IQHE states at $\nu = \pm 2, \pm 6, \pm 10\ldots$ are clearly visible at $B = 4$ T as dips in the longitudinal resistance $R_{xx}$ and plateaus in the Hall resistance $R_{xy}$ traces. Weaker states such as $\pm 4$ and $\pm 8$ are also resolved at $B = 8$ T. The gate voltages where these IQHE states take place in the transport data are in good agreement with those in the MIM data, as plotted in Fig. 1e in the main text. In other words, we do not observe the systematic deviation between transport plateaus and bulk fillings reported by an earlier MIM work[1], which attributes the discrepancy to the presence of a wide edge strip with charge accumulation[2]. It is likely that the proposed effect in that work due to a thick gate dielectric (> 300 nm) is negligible in our work, where the gate dielectric or the bottom hBN is 50 nm in thickness.

# Section S3. MIM signals near the CNP

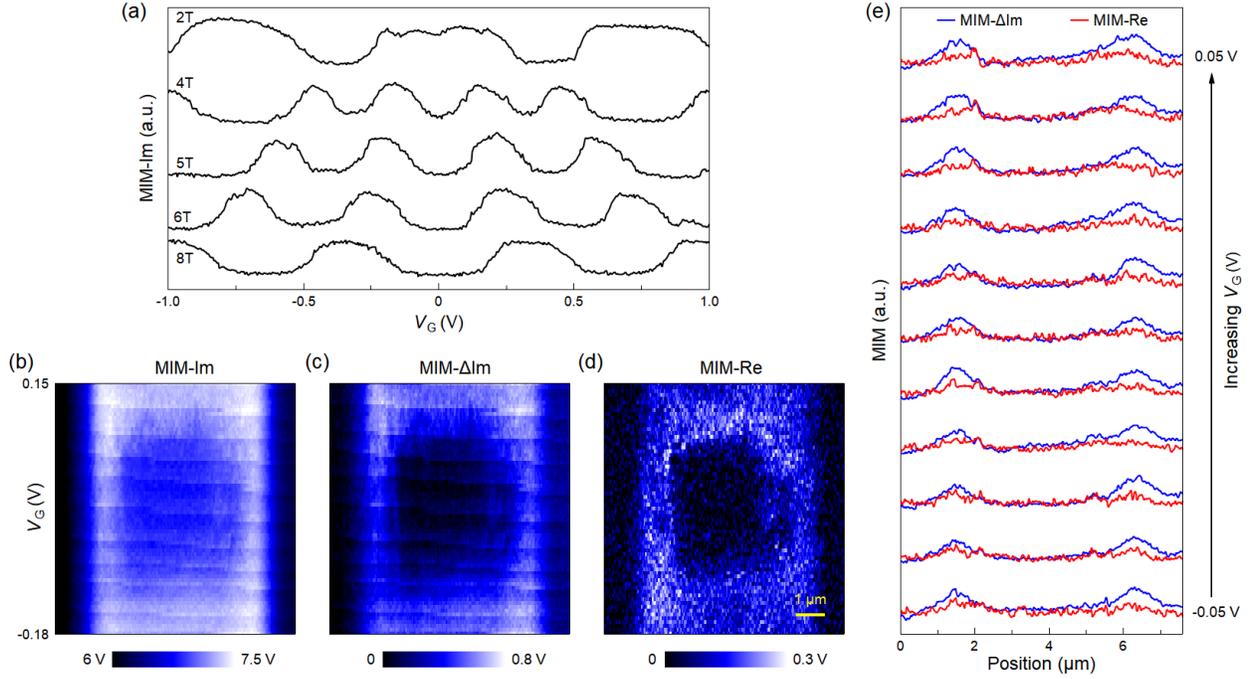

**FIG. S4. (a)** Gate-dependent bulk MIM-Im signals of Device #1 measured under *B*-fields ranging from 2 T to 8 T. The CNP remains at $V_G = 0$ V for all magnetic fields. **(b-d)** Left to right: Gate-dependent MIM-Im, MIM-ΔIm, and MIM-Re images across the sample (dashed line in Fig. 1a in the main text) taken at $B = 4$ T. The glitches are artifacts due to vibration when taking the data. **(e)** MIM-ΔIm and MIM-Re line profiles near the CNP with a step size of 0.01 V in $V_G$. The data are vertically offset for clarity. The conductive edge features persist across the CNP.

The gate voltage corresponding to $\nu = 0$ in Device #1 is determined by gate sweep under various *B*-fields ranging from 2 T to 8 T, as seen in Fig. S4a. In fact, the conductive edge features under intermediate *B*-fields are robust within a narrow range of $V_G$ near the CNP. Figs. S4b-S4d show the MIM line-scan images at $B = 4$ T when $V_G$ sweeps from –0.18 V to 0.15 V. Moreover, the MIM-ΔIm (background-subtracted MIM-Im) and MIM-Re line profiles for $V_G$ from –0.05 V to 0.05 V are plotted in Fig. S4e. It is clear that the peak features near the edge persist across the CNP. In other words, the phenomenon reported in this work is a genuine effect rather than the accumulation of electrons or holes[2] at the edge when $V_G$ is slightly off from the exact CNP.

# Section S4. Raw MIM line profiles at the CNP and $|\nu| \geq 1$ IQHE states

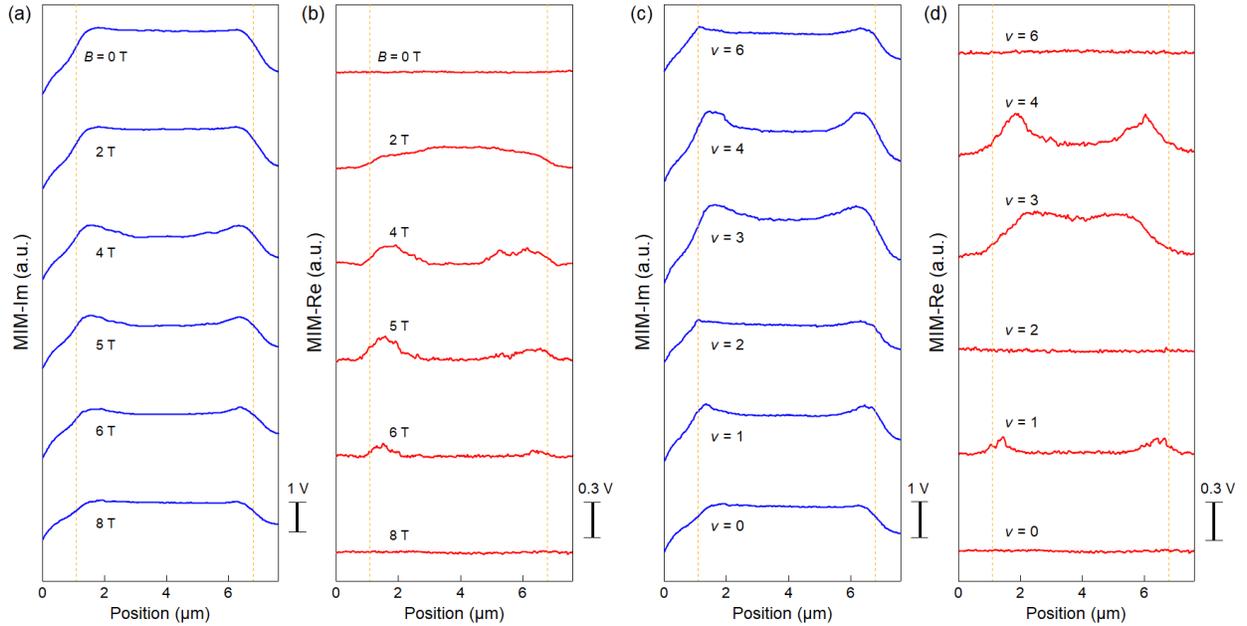

**FIG. S5. (a-b)** MIM-Im and MIM-Re line profiles of Device #1 measured at the CNP under various $B$-fields from 0 T to 8 T. **(c-d)** MIM-Im and MIM-Re line profiles of Device #1 measured at various IQHE states at 8 T. Vertical dashed lines in a–d indicate the etched physical edges.

The MIM-Im (before background removal) and MIM-Re line profiles of Device #1 at the CNP under different magnetic fields are shown in Fig. S5a and S5b, respectively. The slower decrease of MIM signals at the edge than that in the bulk is clearly seen from the raw data. The background subtraction only serves the purpose of highlighting signals near the edge. For completeness, we also show the MIM-Im (before background removal) and MIM-Re line profiles of Device #1 at various IQHE states at 8 T in Fig. S5c and S5d, respectively. The conductive edge features are again obvious from the raw data.

## Section S5. Spatially resolved conductivity at the CNP and $|\nu| \geq 1$ IQHE states

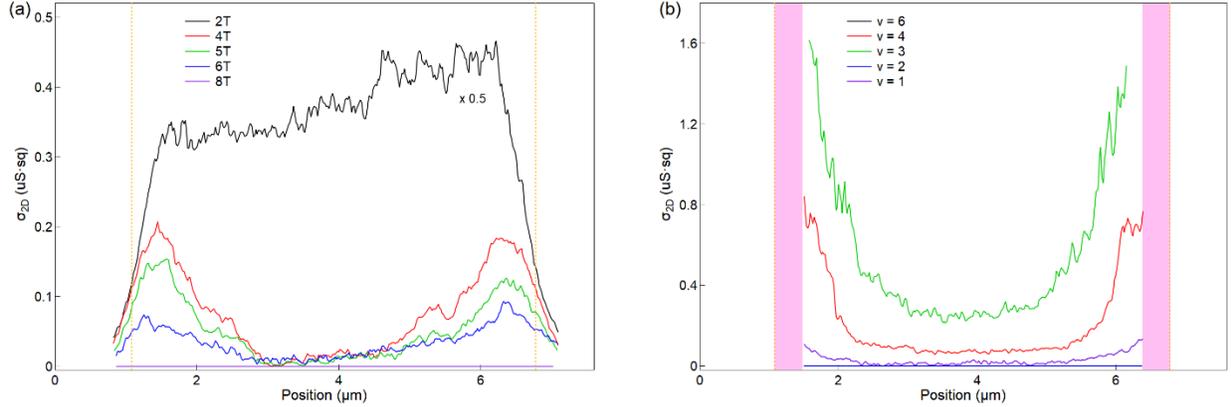

**FIG. S6. (a)** Local conductivity profiles of Device #1 extracted from the MIM data at the CNP under various $B$-fields. **(b)** Local conductivity profiles of Device #1 extracted from the MIM data at various IQHE states at $B = 8$ T. Vertical dashed lines indicate the etched physical edges. The pink vertical regions in **b** correspond to the conductive quantum Hall edge states, where the local conductivity is too high for the scale in this plot.

By comparing the measured MIM data in Fig. 2d and Fig. 3d and the response curves in Fig. 1b, we quantitatively extract the spatially resolved local conductivity profiles at the CNP and $|\nu| \geq 1$ IQHE states of Device #1, as shown in Figs. S6a and S6b, respectively. At the CNP, as $B$ increases, the local 2D conductivity in the bulk decreasing rapidly from above $10^{-3}$ S·sq to below $10^{-9}$ S·sq for $B \geq 4$ T. In contrast, near the graphene edge, the local conductivity drops more gradually with increasing $B$-fields than that in the bulk. For the strong IQHE states such as $\nu = 2$ and 6, the bulk conductivity is below $10^{-9}$ S·sq, whereas the conductivity of the topological edge is beyond our sensitivity window or $\geq 10^{-3}$ S·sq. As the bulk energy gap decreases from $\nu = 1$ to $\nu = 4$ and then $\nu = 3$, the bulk conductivity increases rapidly. The apparent width of the transition region between the conductive edge and the bulk increases as the gap shrinks. The results are in good agreement with the simulated conductivity profiles in Fig. 4 in the main text.

## Section S6. Hartree-Fock calculation of the canted antiferromagnetic (CAF), intervalley coherence (IVC) and charge density wave (CDW) phases

The low-energy physics of monolayer graphene at $\nu = 0$ and under a perpendicular magnetic field is described by the effective Hamiltonian $H$ that can be decomposed into a single-particle term $H_0$, long-range Coulomb interaction $H_C$, and short-range intervalley interactions $H_A$. First, the single-particle Hamiltonian is given by

$$H_0 = \sum_{x_k,\mu} [\epsilon(x_k)\tau_x - \epsilon_Z \sigma_z]_{\mu\mu} c^\dagger_{x_k,\mu} c_{x_k,\mu}$$

, where $x_k$ denotes the guiding center, $\mu = (\tau, \sigma)$ is the composite index for valley and spin, and $c^\dagger_{x_k,\mu}$ and $c_{x_k,\mu}$ are the creation and annihilation operators for an electron in the lowest LL, respectively. The term $\epsilon(x_k)$ represents a smooth edge potential that depends only on the guiding center coordinate. $\epsilon_Z = g\mu_B B_\perp/2$ is the Zeeman coupling to the spin Pauli matrix $\sigma_z$. Second, the long-range Coulomb interaction is given by

$$H_C = \frac{1}{2A} \sum_{\mathbf{q}} V(\mathbf{q}) : \bar{\rho}(\mathbf{q})\bar{\rho}(-\mathbf{q}) :$$

, where $A$ is the sample area, $V(\mathbf{q}) = 2\pi e^2/(\epsilon|\mathbf{q}|)$ is the 2D Fourier transform of the Coulomb potential, and $\bar{\rho}(\mathbf{q})$ is the projected density operator. Note that the Coulomb exchange energy does not depend on the occupancy of a specific flavor and can be disregarded when comparing the energy of ground states. Finally, the short-range anisotropic interaction energy is given by

$$H_A = \frac{1}{2A} \sum_{\alpha=x,y,z} g_\alpha \sum_{\mathbf{q}} : \bar{T}^\alpha(\mathbf{q}) \bar{T}^\alpha(-\mathbf{q}) :$$

, where $g_{\alpha=x,y,z}$ are phenomenological intervalley coupling constants and $\bar{T}^\alpha(\mathbf{q})$ the projected valley-isospin operators. Following the analysis in Ref. [3], for negative $g_x = g_y = g_\perp < 0$ and $|g_\perp| < g_z$, the system favors an antiferromagnetic (AF) ordering with spontaneous staggered alignment of isospins in the valley $xy$-plane. For $g_\perp < 0$ and $|g_\perp| > g_z$, the IVC phase has the lowest energy. For $g_z < 0$ and $|g_\perp| < |g_z|$, the CDW phase is the ground-state configuration. Note that the analysis above is in the absence of a magnetic field. A finite Zeeman term seeks to

align spins along the z-axis. For the AF ordering, the system will compromise by forming the CAF phase, where the antiparallel isospins develop a tilt out of the xy-plane.

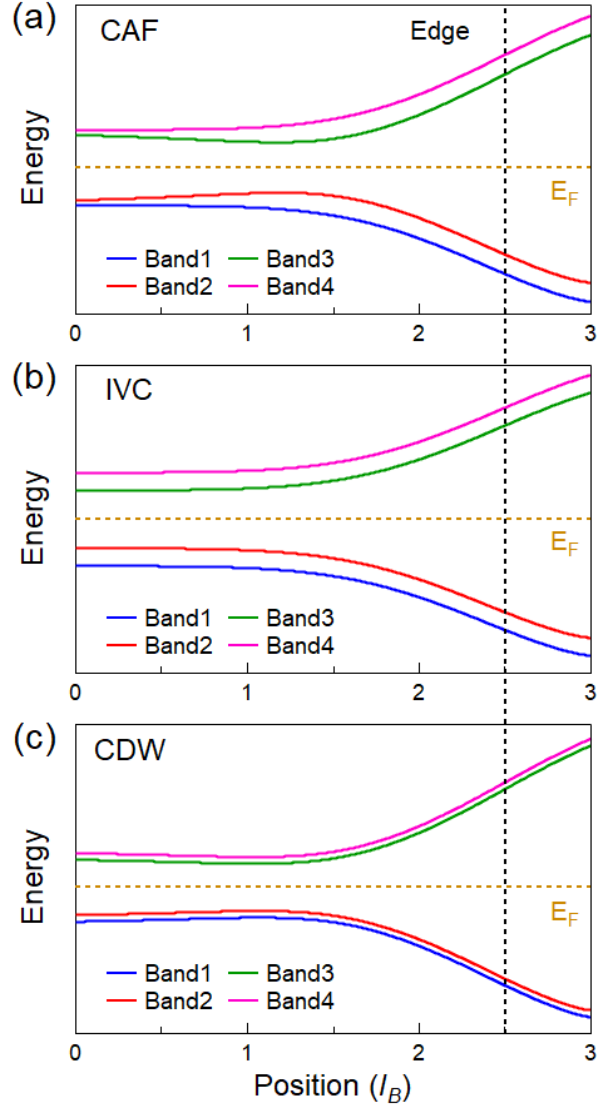

**FIG. S7.** Calculated Hartree-Fock energy bands of (a) CAF, (b) IVC, and (c) CDW phases as a function of the guiding center coordinate $x_k$ at $\nu = 0$. The edge position is at $x_0 = 2.5 l_B$, indicated by the vertical dashed line. The position is in unit of $l_B$.

To model a finite sample geometry, we introduce a hard-wall confining potential along one edge, described by a step function $U(x_0) = U_0\, \Theta(x - x_0)$, where $x_0$ marks the sample edge. When projected onto the $N = 0$ LL, this sharp potential is experienced by the electrons as a smooth diagonal potential in the guiding center basis, $V(x_k) = \frac{U_0}{2}\operatorname{erfc}\left(\frac{x_0 - x_k}{l_B}\right)$. This confining potential is added to the single-particle term of our Hartree-Fock Hamiltonian. For CAF, IVC, and CDW phases, the calculated Hartree-Fock energy spectra for the four flavors at the CNP are plotted in Figure S7a – S7c. The band structures reveal two key features. First, the energy gap between occupied and unoccupied states remains finite across the entire sample for all three phases. The absence of a band crossing at the Fermi level signifies that, for these quantum Hall ferromagnetism phases, there are no gapless single-particle edge states. Second, when approaching the graphene edge, the ν = 0 gap between the highest occupied band (band 2) and unoccupied band (band 3) increases monotonically for the CDW and IVC phases but changes non-monotonically for the CAF phase. Moreover, the width of the gap-narrowing region in the CAF phase is on the order of several magnetic lengths. These features are consistent with previous calculations[3-5] and the spatial evolution of the local conductivity indicated in Fig. 2. We use this result for the finite-element analysis (FEA)[6] simulation of CNP states in the following sections.

We now compute the local conductivity of neutral graphene under various conditions. What is relevant to our experiment is the temperature-dependent conductivity at microwave frequencies, which are essentially *dc*. This quantity is contributed by thermally activated electrons in the empty flavor states and by thermally activated holes in the occupied flavor states. The density of these carriers follows an activated temperature dependence, i.e., $n \propto e^{-\frac{\Delta}{2k_BT}}$. Thus, the local conductivity can be expressed as:

$$\sigma_{xx}(\mathbf{r}) = \sum_i A_i e^{-\frac{\Delta_i(\mathbf{r})}{2k_BT}}$$

, where *i* is the index for bands with different spins and valleys, $A_i$ is a constant for each band (assumed to be the same for all bands), $\Delta_i(\mathbf{r})$ is the local gap of band *i* at position $\mathbf{r}$, $k_B$ is the Boltzmann constant, and *T* is the temperature. The calculated local conductivity profiles across the graphene edge for CAF, IVC, and CDW phases under various magnetic fields are shown in Fig. S8. The relevant magnetic fields here are on the order of 2 – 8 T, consistent with the MIM experiment. After considering the gate screening and inter-band screening, the thermal activation energy in our calculation corresponds to a temperature scale of 1 K, again in agreement with the experiment. For the CAF phase, the calculated conductivity increases when approaching the graphene edge, reaches a maximum, and then decreases. In contrast, for the IVC and CAF phases, the calculated conductivity does not show an appreciable enhancement at the edge. In summary, by comparing the experimentally measured local conductivity profiles in Fig. S6a and theoretically calculated values in Fig. S8, we conclude that the ground state of graphene indeed corresponds to the CAF phase.

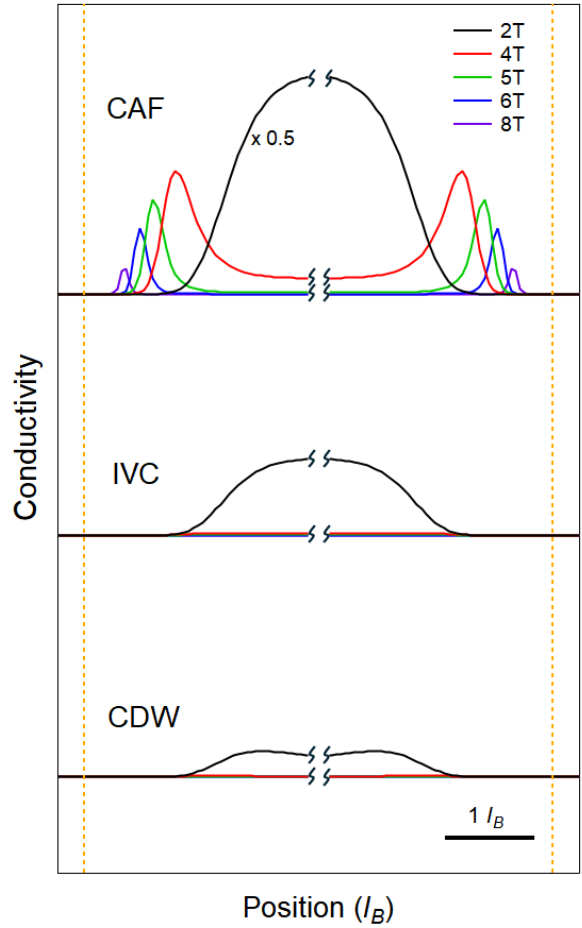

**FIG. S8.** Calculated conductivity profiles based on the Hartree-Fork energy bands of CAF, IVC and CDW phases as a function of the guiding center coordinate $x_k$ at $\nu = 0$ under various magnetic fields from 2 T to 8 T. The vertical dashed lines indicate the etched physical edges. The position is in unit of $l_B$, as labeled in the plot. Note that the calculation still shows a very small edge-like conductivity peak at 8 T for the CAF phase, which could be suppressed by disorders in the device in our experiment.

## Section S7. FEA simulation of the CNP and IQHE states

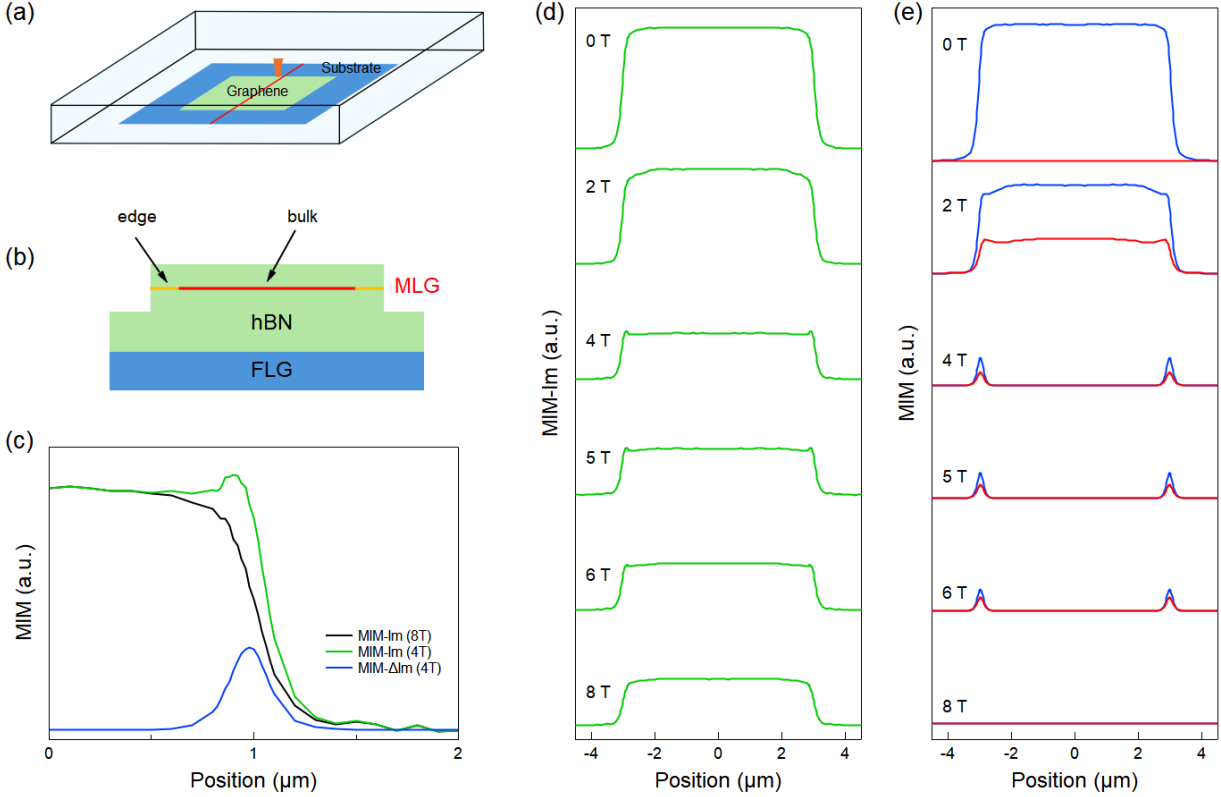

**FIG. S9. (a)** 3D FEA model for the simulation of MIM signals at the CNP. **(b)** Schematic of the cross-sectional view along the red line in **a**. **(c)** Simulated MIM-Im line profiles at the CNP with $B = 4$ T and 8 T, as well as the background subtraction to obtain MIM-$\Delta$Im. **(d)** Simulated MIM-Im and **(e)** MIM-$\Delta$Im/-Re line profiles with various $\sigma_b$ and $\sigma_e$ (see Table S1) to qualitatively reproduce the experimental data at the CNP.

We performed FEA[6] to simulate the MIM signals at the CNP (Fig. S9) and IQHEs (Fig. S10) by commercial software COMSOL 6.2. For both models, the entire 3D simulated volume is 38 μm (length) × 32 μm (width) × 5 μm (height). The lateral size of the substrate (50 nm bottom hBN / 10 nm FLG) region is 28 μm (length) × 18 μm (width). The lateral size of the active sample (10 nm top hBN / graphene / 10 nm bottom hBN) region is 18 μm (length) × 6 μm (width). The parameters are chosen to be as close to the actual dimensions as possible. As it is very difficult to create mesh for the 0.34 nm-thick graphene, we use $t = 5$ nm as the thickness in the simulation and quote $\sigma_{2D} = \sigma_{3D} \cdot t$ when presenting the results. The MIM tip has a diameter of 300 nm at the apex and a cone-shaped body with a height of 1 μm. The tip-sample distance is set to be 90 nm.

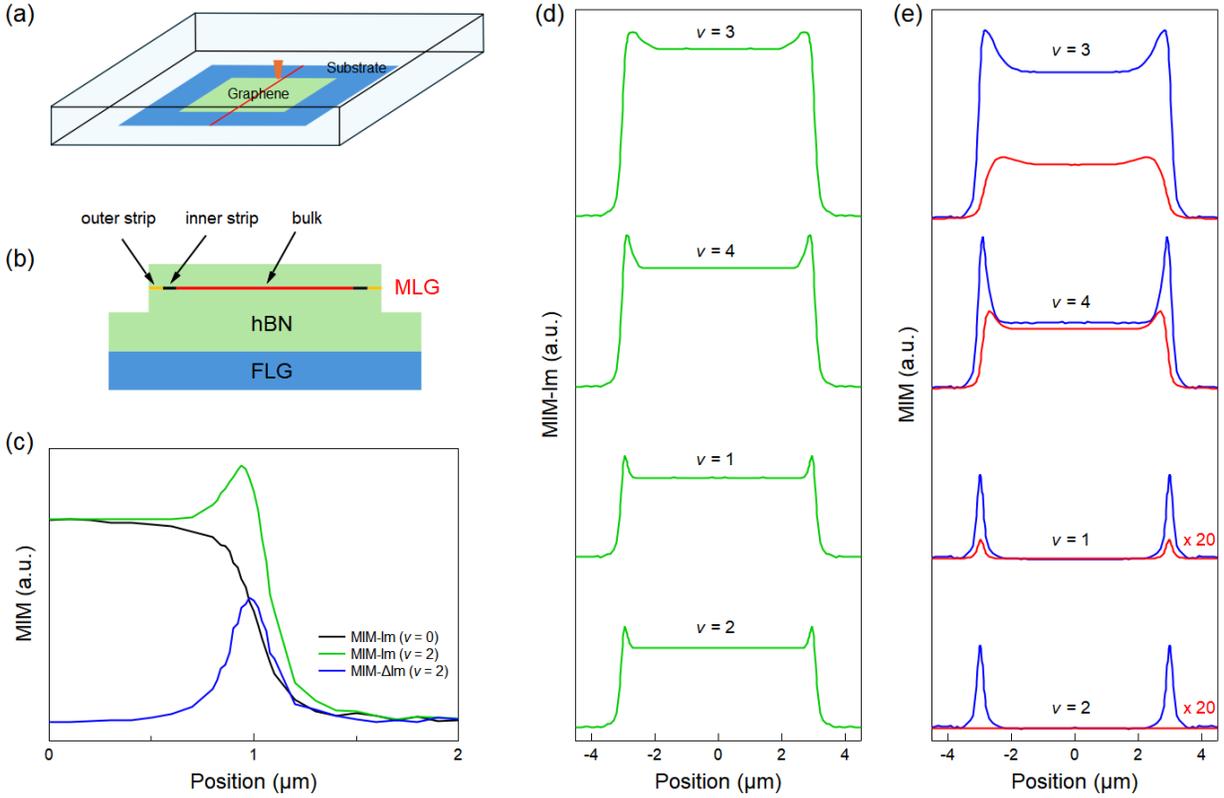

**FIG. S10. (a)** 3D FEA model for the simulation of MIM signals at the IQHE states. **(b)** Schematic of the cross-sectional view along the red line in **a**. **(c)** Simulated MIM-Im line profiles at ν = 0 and ν = 2, as well as the background subtraction to obtain MIM-ΔIm. **(d)** Simulated MIM-Im and **(e)** MIM-ΔIm/-Re line profiles with various $\sigma_b$ and $\sigma_e$ (see Table S2) to qualitatively reproduce the experimental data at various IQHE states. Note that the MIM-Re line profiles at ν = 1 and ν = 2 are amplified by a factor of 20 to highlight small features.

The only difference between the two models is the edge structure. The CNP model in Fig. S9 contains a single edge strip with 30 nm in width around the perimeter of graphene. Note that the estimated width in our simulation, corresponding to a few magnetic lengths ($l_B \approx 10$ nm at 8 T), is based on the calculated Hartree-Fork bands near the edges (Section S6). The bulk sheet conductance ($\sigma_b$) and edge conductance ($\sigma_e$) for qualitatively replicating the MIM data under various $B$-fields are summarized in Table S1. Note that $\sigma_b$ and $\sigma_e$ values on the order of 0.01 ~ 0.1 $e^2/h$ are physically reasonable due to thermal activation across the gap. The background subtraction is illustrated in Fig. S9c, where the simulated MIM-ΔIm signals by removing the $B = 8$ T background indeed exhibit a peak at the edge. Finally, the simulated MIM line profiles at various fields are plotted in Figs. S9d and S9e, which closely resemble the measured data in Fig. 2d in the main text.

Table S1: 2D sheet conductance in the CNP model

| $B$ (T) | $\sigma_b$ ($e^2/h$) | $\sigma_e$ ($e^2/h$) |
|---|---|---|
| 0 | $10^3$ | $10^3$ |
| 2 | 0.05 | 0.25 |
| 4 | 0 | 0.2 |
| 5 | 0 | 0.15 |
| 6 | 0 | 0.01 |
| 8 | 0 | 0 |

The IQHE model in Fig. S10 contains two edge strips around the perimeter of the graphene layer. The outer and inner strips represent the compressible and incompressible edge states, respectively. The widths of both strips are assumed to be 10 nm, which is close to the magnetic length at 8 T. The outer chiral edge is assumed to be perfectly conductive with $\sigma_{xx} = \infty$ (5 S·sq in the simulation). The bulk has an off-diagonal 2D conductivity of $\sigma_{xy} = -\sigma_{yx} = \nu e^2/h$, where $e$ is the elementary charge and $h$ is the Planck constant. The diagonal 2D conductivities of the bulk ($\sigma_b$) and inner edge ($\sigma_e$) regions for qualitatively replicating the MIM data at various filling factors are summarized in Table S2. The background subtraction is again illustrated in Fig. S10c, where the simulated MIM-$\Delta$Im signals at $\nu = 2$ by removing the $\nu = 0$ background indeed exhibit a peak at the edge. Finally, the simulated MIM line profiles at various filling factors are plotted in Figs. S10d and S10e, which closely resemble the measured data in Fig. 3d in the main text.

Table S2: 2D sheet conductance in the IQHE model

| Filling factor $\nu$ | $\sigma_b$ ($e^2/h$) | $\sigma_e$ ($e^2/h$) |
|---|---|---|
| 3 | 0.05 | 0.05 |
| 4 | 0.01 | 0.01 |
| 1 | 0 | 0.001 |
| 2 | 0 | 0 |